\documentstyle[preprint,revtex]{aps}
\begin{document}
\draft
\preprint{Submitted to  PRA, 25 June 1992 }
\begin{title}

  Dynamic Scaling  of Coupled Nonequilibrium Interfaces
\end{title}

\author{Albert-L\'aszl\'o Barab\'asi$^{*}$ }
\begin{instit}

 Center for Polymer Studies and Department of Physics

 Boston University, Boston, MA 02215 USA
\end{instit}
\begin{abstract}

  We propose a simple discrete model to  study the
nonequilibrium fluctuations of two locally coupled 1+1 dimensional
systems (interfaces).
 Measuring numerically  the
tilt-dependent velocity  we  construct a set of stochastic
 continuum
equations describing the fluctuations in the model. The
scaling predicted by the equations are
 studied analytically using dynamic renormalization group and compared
with simulation results. When the coupling is symmetric, the well known
KPZ exponents are recovered.
If one
of the systems is fluctuating independently, an increase in the
roughness exponent is observed for the other one.

\end{abstract}
\pacs{PACS numbers: }

The dynamics of equilibrium and nonequilibrium interfaces has attracted
much attention recently \cite{[1]}. The continuum theory proposed
by Kardar, Parisi and Zhang (KPZ) \cite{[2]} provided a very successful
analytic
approach for different problems, ranging from spin systems to growth
models.
However usually the dynamics of interest is dominated  by a strong coupling
fixed point, making the perturbative approach inconclusive.
 In such a situation simple discrete models allow us to obtain
detailed information on the dynamic scaling.

	In this paper   a generalized single step  model is introduced
to study the nonequilibrium dynamics of two coupled  $1+1$ dimensional systems.
Investigating the tilt dependence of the growth velocity we find that
the set of nonlinear equations describing the fluctuations of the
  two surfaces  $h_0(x,t)$ and $h_1(x,t)$ is
\begin{mathletters}
\begin{equation}
 \partial_t h_0 = \nu_0 \partial^2_x h_0 +\lambda_0 (\partial_x h_0)^2 +
 \gamma_0 \partial_x h_0 \partial_x h_1 + \varphi_0 (\partial_x h_1)^2 +
\eta_0
\end{equation}
\begin{equation}
\partial_t h_1 = \nu_1 \partial^2_x h_1 +\lambda_1 (\partial_x h_1)^2 +
\gamma_1 \partial_x h_0 \partial_x h_1 + \varphi_1 (\partial_x h_0)^2 +
\eta_1
\label{eq:(1)}
\end{equation}
\end{mathletters}
 where $\partial_t$ and $\partial_x$ denote the partial derivatives
with respect to $t$ and $x$. The noise is assumed to have zero mean
and uncorrelated
 $\langle\eta_i(x,t) \eta_i(x',t')\rangle = D_i\delta(t-t')\delta(x-x')$
with $i=0,1$.
The scaling of the fluctuations are determined numerically from the
discrete model and the results are compared with the predictions of the
dynamic renormalization group (DRG) calculations. A slightly modified
version of the model allows us to study the evolution of a growing
interface perturbed by a nonequilibrium field.  The increase of the
roughness exponent of the surface observed numerically is supported by
the analytic predictions.

Equation (1) represents the most general form describing the
fluctuations of two {\it coupled nonequilibrium } systems with {\it local
} dynamics.
It describes the evolution of two interfaces which are moving with the
same mean  velocity in an inhomogeneous media and are coupled by some
local interaction. More generally it describes the growth of an
interface $h_0(x,t)$ in the presence of a nonequilibrium field $h_1(x,t)$,
assuming that the interface and the field are strongly
affecting each others behavior. In this respect it might provide a
suitable analytic  framework with which  to describe the recent surface growth
experiments in which
the control of other physical factors was not possible.   Such a
perturbing field might be the density or pressure of the fluid during
fluid displacement in porous media \cite{[3]}, the concentration of the
nutrient
in bacterial growth \cite{[4]}, or the density of the mesoscopic particles and
impurities in the imbibition experiments \cite{[5]}.

On the other hand the study of equation (1) is of general interest in
the context of the recent efforts to understand the general properties of
the  nonequilibrium stochastic systems.
Recently Ertas and Kardar (EK),  studying the motion of a single
flux line in random environment,  have presented detailed
investigation of  (1) for
 $\gamma_0=\lambda_1=\varphi_1=0$ \cite{[6]}.  Imposing an additional
condition,
$\varphi_0=0$, the problem reduces to the convection of a passive scalar
($T \equiv h_1(x,t)$) in a random velocity field  ($\vec v \equiv \vec \nabla
h_0$) \cite{[7]}. For $\gamma_1=\varphi_1=0$ equation (1)
describes the fluctuations of an interface $h_0(x,t)$ evolving  under the
influence of an {\it independent}  nonequilibrium  field $h_1(x,t)$.\cite{[8]}

{\it  The Model -- } Our goal is to introduce a simple model to study the
scaling of the two coupled interfaces. For this  we  use the  mapping
between the single-step model and the  driven hard-core lattice-gas
on a one-dimensional chain \cite{[9]}. In the mapping each step
along the surface $h_0(x,t)$  is associated with a site on
the chain $H_0$. A site $j$ is said
to be occupied (i.e.  $H_0(j)=1$) if we have an upward step on
the interface $h_0(x=j,t)$, and
is empty  ($H_0(j)=0$) otherwise. Two one-dimensional lattices, $H_0(j)$
and $H_1(j)$  with ($j=1,L$),  are
filled with probability ${1 \over 2}$. First  choose randomly a site $j$ on
one of the lattices (for example $H_0$). If
it is filled (i.e. $H_0(j)=1$), the quantity
$\vartheta =[H_0(j-1)+H_0(j)+H_0(j+1)-1.5][H_1(j-1)+H_1(j)+H_1(j+1)-1.5]$
determines  the
direction in which the chosen particle wants to move. If $\vartheta > 0$ the
$H_0(j)$
particle will try to move to the left, otherwise to the right. The
step can be completed only if the neighboring site in the chosen
direction is empty.  Then  choose randomly an another particle on the other
lattice and repeat the same procedure. The unit of time is defined as  $L$
trials per lattice, where $L$ is the system size.
Periodic boundary conditions are imposed on both of the lattices.

    The quantity $[H_0(j-1)+H_0(j)+H_0(j+1)]$ is related to the   local slope
of
the  interface $h_0(x=j,t)$, thus with a positive
$\vartheta$  the two interfaces have the same slope and is generating
a growth on the site $j$, while a negative $\vartheta$ corresponds to different
slopes and   results in  a decrease
of the height.

 We are interested in the scaling
of the fluctuations  characterized by the dynamic scaling
of the width \cite{[1]}
\begin{equation}
w^2_i(L,t) = < (h_i(x,t)-\bar h_i(t))^2> = L^{2 \chi_i} f({t \over
 L^{z_i}})
\label{eq:(2)}
\end{equation}

where $\chi_i$ is the roughness exponent for the interface $h_i(x,t)$, and the
dynamic exponent $z_i$ describes the scaling of the relaxation times with
length; $\bar h(t)$ is the mean height of the interface at the moment
$t$ and  the $<...>$ symbols denote ensemble and space average.
  The scaling function $f_i$   has the properties $f_i(u \rightarrow 0) \sim
u^{2 \beta_i}$ and $f_i(u \rightarrow \infty) \sim const_i$, where
$\beta_i = z_i / \chi_i$.

 {\it Tilt-dependence --} In order to identify the relevant nonlinear
terms determining the scaling behavior in the model we have measured
the tilt dependent velocity \cite{[10]}, imposing a global slope on the
interfaces
$h_0(x,t)$ and $h_1(x,t)$.
A positive  slope is induced on $h_0(x,t)$ by increasing the
concentration of the $H_0$ particles, a negative slope  is a result of a
 decrease
in  concentration. Fig. 1 shows the tilt-dependent velocity $v_0$
and $v_1$, where
$v_i=\sum_{j=1,L} \langle
\partial_t h_i(x=j,t) \rangle$.
The velocities which may explain the observed behavior and are compatible
with the $x \rightarrow -x$ symmetry of the model have the form:
$v_0 = k-a_0(\partial_x h_0 \partial_x h_1)
-b_0(\partial_x h_0)^2$ and     $v_1= k+a_1(\partial_xh_0 \partial_x h_1) - b_1
(\partial_x h_0)^2$, where $k, a_i, b_i$ are positive constants.
 The tilt in $h_0(x,t)$ induces a
negative {\it  mixing } term  ($\partial_xh_0 \partial_x h_1$) in $v_0$
and a positive one in $v_1$. But since in
 the model  the
influence of the $h_0(x,t)$ to $h_1(x,t)$ is identical to the influence of
$h_1(x,t)$ to $h_0(x,t)$
(symmetric coupling), without tilt one expects the same sign for both mixing
terms.
The symmetric
coupling is  responsible for the presence of
a $\varphi_0(\partial_x h_1)^2$ term in $v_0$ and a $\lambda_1(\partial_x
h_1)^2$ term in
$v_1$, which can be observed  if the tilt is in $h_1(x,t)$ instead of
$h_0(x,t)$.
Based on
these measurements we conclude that equation (1) contains all the relevant
nonlinear
terms  determining the scaling of the two interfaces, and it describes  the
proposed model if $\lambda_i < 0$ and
$\varphi_i < 0$.
 Higher order terms might
 be responsible for the unusual behavior observed for large tilts,
but they are in fact irrelevant concerning the scaling.

 {\it Dynamic Renormalization Group  Analysis --} The scaling behavior of
the equation  (1)   in general can be investigated using one-loop DRG
\cite{[7],[11]}.  Rescaling the
parameters   $x \to e^\ell x$, $t \to e^{\ell z} t$, and
 $h_i \to  e^{\ell \chi_i }h_i$, we obtain the following flow equations for
the coefficients:
\begin{eqnarray}
  { d \nu_0 \over d \ell} = \nu_0 \lbrack z - 2 +   { K_1 \over \nu_0 } (
{\lambda_0^2 D_0 \over \nu_0^2} + {\gamma_1 \varphi_0 D_1 \over 2 \nu_1^2}) +
{K_1 \gamma_0 \over 2 \nu_0 (\nu_0 + \nu_1)} ( {2 \varphi_1 D_0 \over
\nu_0} + {\gamma_0 D_1 \over \nu_1 } ) \nonumber\\
+ {K_1 \gamma_0 \over \nu_0 } {
\nu_0 -\nu_1 \over (\nu_0 + \nu_1)^2 } ( { 2 \varphi_1 D_0 \over \nu_0 } -
{ \gamma_0 D_1 \over \nu_1 }) \rbrack \nonumber
\end{eqnarray}
\begin{eqnarray}
 { d D_0 \over d \ell }= D_0  \lbrack z-2\chi_0 - 1 +{\lambda_0^2
D_0 K_1 \over \nu_0^3 } +{ \gamma_0^2 D_1 K_1 \over \nu_0 \nu_1
(\nu_0+\nu_1)} \rbrack + {\varphi_0^2 D_1^2 K_1 \over \nu_1^3 }
\label{eq:(3)}
\end{eqnarray}
\begin{eqnarray}
{ d \lambda_0  \over d \ell } =  \lambda_0 \lbrack z+\chi_0-2
\rbrack + {K_1 \over \nu_0 + \nu_1 } ( { 2 \varphi_1 D_0  \over \nu_0 } -
{ \gamma_0 D_1 \over \nu_1 } ) ( {\gamma_0 \lambda_0 \over \nu_0}{\nu_0
- \nu_1 \over \nu_0 + \nu_1 } + { 2 \varphi_0 \varphi_1 \over \nu_1 } - {
\gamma_0 \gamma_1 \over \nu_0 + \nu_1 } ) \nonumber
\end{eqnarray}
\begin{eqnarray}
 { d \gamma_0 \over d \ell } = \gamma_0 \lbrack z + \chi_1 - 2
\rbrack & + &
{K_1 \over \nu_0 + \nu_1 } \lbrack ( { 2 \varphi_1 D_0 \over \nu_0
} - {  \gamma_0  D_1 \over \nu_1}) (   \gamma_0 { (\gamma_0 - 2 \lambda_1)
\over  \nu_0 + \nu_1 } - { \lambda_0 \varphi_0
\over  \nu_0 }
 + { \gamma_1 \varphi_0  \over 2 \nu_1 } )  + \nonumber\\
&& ({ 2 \varphi_0 D_1 \over \nu_1 }  - { \gamma_1 D_0 \over \nu_0 })
(  \gamma_0 { \gamma_1 - 2 \lambda_0 \over \nu_0 + \nu_1} +  { \lambda_0
\gamma_0
  \over 2  \nu_0  } - { \varphi_0 \varphi_1 \over \nu_1 }   ) \rbrack \nonumber
\end{eqnarray}
\begin{eqnarray}
{ d \varphi_0 \over d \ell } = \varphi_0 \lbrack 2 \chi_1-\chi_0+z-2
\rbrack  +
{ K_1\over \nu_0 + \nu_1 }(  {\gamma_1 D_0 \over \nu_0} - { 2 \varphi_0
D_1 \over  \nu_1} ) ( \varphi_0 { \gamma_1 - 2 \lambda_0 \over \nu_0 } +
\gamma_0
{ \gamma_0 - 2 \lambda_1 \over \nu_0 + \nu_1 })\nonumber
\end{eqnarray}
The flow equations for the other five missing coefficients  can be obtained
from
(\ref{eq:(3)}) replacing $\nu_0
\leftrightarrow \nu_1$, $D_0 \leftrightarrow D_1$, $\lambda_0 \leftrightarrow
\lambda_1$, $\gamma_0 \leftrightarrow
\gamma_1$, $\varphi_0 \leftrightarrow \varphi_1$ and $\chi_0 \leftrightarrow
\chi_1$.

Obtaining the exponents  from (\ref{eq:(3)})  is not straigforward because
of the large number of parameters involved, but
important  results can be obtained by  making use of the nonperturbative
properties of (1) combined with the direct integration of (\ref{eq:(3)}).

	If $2 \varphi_0 \nu_0 D_1 = \gamma_1 \nu_1 D_0$ and
 $2 \varphi_1 \nu_1 D_0 = \gamma_0 \nu_0 D_1$ (fluctuation-dissipation
(FD) subspace) the joint probability \cite{[6]}
\begin{eqnarray*}
 \wp [h_0(x,t), h_1(x,t)]= exp( - \int dx [{\nu_0 \over 2 D_0} (\partial_x
h_0)^2 +
{\nu_1 \over 2 D_1 } (\partial_x h_1 )^2 ] )
\end{eqnarray*}
 is a solution of the Fokker-Planck equation following from (1). This
provides us the exact exponents $\chi_i=1/2$.
The  nonlinear
terms do not renormalize, resulting in the scaling relation $z + \chi_i
= 2$.
 Direct integration of (\ref{eq:(3)})  shows that for positive
$\alpha_i, \varphi_i$ and $\gamma_i$ the flow converges to the FD
subspace (see Fig 2), thus resulting in the superdiffusive  exponents
 $\chi_i= 1/2$ and $z=3/2$. A  change of variables $h_0 \to -h_0$ and $h_1 \to
-h_1$ indicate the same exponents if {\it  all} the coefficients of the
nonlinear terms are negative.

As the  velocity measurements indicated,  $\lambda_i < 0$ and
 $ \varphi_i < 0 $ in the model. If $\gamma_i$ is also negative,
the FD subspace dominates the behavior.
The DRG
does not provide exact results in the $\gamma_i >0$  case;
 direct integration of (\ref{eq:(3)})   indicates a strong coupling fixed
point, with
diverging $\lambda_i$, $\gamma_i$ and $\varphi_i$.

 Fig. 3 shows the scaling of the
 width $w_0^2(L,t)  \sim
t^{2 \beta_i} $.
The exponents were
determined using saturated systems of size $L=50,100,200,400$  and
unsaturated systems of size $L=1000, 2000, 4000$.
The roughness exponent $\chi_i$  was obtained from
 the best collapse for the
time dependent width (\ref{eq:(2)}).
 All the simulations indicated  superdiffusive exponents,
 giving as result  $\chi_i = 0.52 \pm 0.03$ and
 $\beta_i=0.32 \pm 0.02$.
Concluding this section we note that the simulations are in perfect
agreement with the predictions of the DRG for $\gamma_i < 0$, so it is
very likely that this is the sign of the mixing term in the model. But
since no analytical results are available for $\gamma_i > 0$, we can not
rule out the possibility that for this sign the fluctuations of the two systems
are also
characterized by the superdiffusive exponents.

In the following two sections we study a variant of the proposed model,
in which one of the interfaces fluctuates independently.  This
corresponds to the above mentioned situation, when an interface
$h_0(x,t)$ grows in the presence of an equilibrium or  nonequilibrium
field $h_1(x,t)$. The fluctuations of the  field is assumed to be
independent of the  growing interface. The presence of
the field results in an enhancement of the roughness of the growing
surface.

{\it  The  $\lambda_0 = \varphi_0 = \varphi_1 = \gamma_1 = 0$ case -- }
 The $h_1$ interface  fluctuates independently of $h_0$
in the following version of the model: The randomly chosen particle on the
$H_0$
chain will try  to move in the direction  determined by the sign of
 $\vartheta$, but  the
one chosen on $H_1$ will try to move left with a probability $p$ and right with
probability $1-p$, independently of $\vartheta$. On $H_1$ we have the
 single step model with
evaporation.    If $p \not= 1/2$, the fluctuations of the $h_2$
interface are  described by the KPZ equation, thus $\gamma_1 = \varphi_1 =
0$ in (1).
For $p=1$  measurements  on the tilt dependence of
$v_0$ indicate the absence of the $\lambda_0$ and $\varphi_0$ terms: the
velocity does not change if we impose a tilt either on $h_0$ or on
$h_1$, the only influence is the decrease of $v_1$ for a tilt on
$h_2$, as a result of the nonvanishing negative $\lambda_1$ term. But these
results  do
not indicate the absence of the $\gamma_0$ term: the fact that it has no
influence on the velocity may come from the vanishing contribution of
 $< \partial_x h_0 \partial_x h_1 >$ , which is acceptable
 considering that $h_1$ fluctuates independently
of $h_0$.
The exponents determined from the model are
 $\chi_0=0.64 \pm 0.03$ and $\beta_0 = 0.33 \pm 0.02 $, together
with the known exponents of the KPZ
equation: $\chi_1 = 1/2$ and $\beta_1 = 1/3$ (see Fig 4). These exponents are
in
good agreement  with
those obtained by EK from the direct integration of the equations
 (1)
for $\lambda_1 < 0$ and $\gamma_0 > 0$ \cite{[13]}.
  The flow equations  (\ref{eq:(3)}) show that $\lambda_1$ scales to zero, thus
the
 DRG is not conclusive for these coefficients. It indicates
only that $z_0 > z_1$, which agrees with the numerical findings. A
support for the positivity of $\gamma_0$ comes from the DRG result for
$\gamma_0 < 0$, the one-loop exponents being  $\chi_0 = 3/4$ and
$\beta_0 = 1/2$, considerably larger than those observed numerically in
the model \cite{[14]}.

{\it  The  $\lambda_0 = \lambda_1 =\varphi_0=\varphi_1 = \gamma_1 = 0$
case -- }
If in the second variant of the model we choose $p =
1/2$, the $h_1$ interface fluctuates in equilibrium with
$\lambda_1=\varphi_1 = \gamma_1$,
leading to the Edwards-Wilkinson (EW) exponents $\chi_1=1/2$ and
$\beta_1=1/4$ \cite{[15]}.
The situation
in the model is the following: the $h_1$ interface fluctuating  in
equilibrium influence the fluctuation of the $h_0$ interface
through the mixing term $(\partial_x h_0 \partial_x h_1)$.
 The velocity measurements again
confirm the absence of the
$\varphi_i$, $\lambda_i$ terms.
 The simulations  indicate the nontrivial exponents $\chi_0=0.68 \pm 0.02$ and
$\beta_0=
0.34\pm 0.02 $, together with the EW exponents for $h_1$. The DRG in the
present form can not be applied
  because of the different scaling of the time in the two interfaces
(i.e. $z_0 \not= z_1$).

{ \it Discussion -- } In principle the existence of a term
$(\partial_x^2 h_1)$ in (1a) (and
$(\partial_x^2 h_0)$ in (1b)) can not be excluded neither  from symmetry
considerations nor from velocity measurements. These terms can be
transformed away at the expense of a cross-correlated noise in (1) (i.e.
the new  noise has the form
 $\eta_0=\zeta_0 + a \zeta_1$ and $\eta_1=\zeta_0 - a \zeta_1$, where $a$
is a constant and $\zeta_0$ and $\zeta_1$ are uncorrelated noises).
 Further study is necessary to determine the
influence of a such a noise on the scaling and on the DRG flow.

 The  results presented below are
just a small subset of the possible behavior following from  (1) and
(\ref{eq:(3)}). Further investigations are necessary to understand the scaling
for
other choices of the coefficients. In this paper we have focused mostly
 on those parameter values which were accessible
through the studied discrete  models.  Determining the relevant
nonlinear terms using tilt-dependent velocity measurements, we have shown
that the model is described by  two coupled
nonlinear Langevin equations of form (1).
For symmetric coupling the model allowed us
to determine the scaling exponents  which were consistent with the exact
results obtained from the DRG.
The study of two particular cases were also possible using a modified
version of the model. In the first one a nonequilibrium field $h_0$ was
influenced by $h_1$, whose oscillations are described by the KPZ
equations, while in  the second case $h_1$ was fluctuating in equilibrium. Both
cases leaded  to considerable  enhancement of the roughness exponent
$\chi_0$.

We thank S.V. Buldyrev, D. Ertas, G. Huber, T. Hwa, J. Kert\'esz, T. Vicsek,
and H.E.
Stanley for useful discussions and comments on the manuscript.
  This research was partially founded by the Hungary-USA exchange
program of the Hungarian Academy of Sciences and National Science Foundation.
The Center for Polymer
Studies is supported by NSF.

\figure{ The tilt dependent velocity $v_0$ (empty
symbols) and $v_1$ (filled symbols) for system sizes $L=125$ (circles),
$L=250$ (triangles) and $L=500$ (squares). Each set of data points was
obtained by averaging over  250,100 and 25 independent runs.
 The inset shows the simplest polynomial in the tilt
$(\partial_x h_0)$ of the form $v_0 = 0.031 - 0.34 (\partial_x h_0)
-0.12(\partial_x h_0)^2$ and     $v_1= 0.031+0.02 (\partial_xh_0)  - 0.02
(\partial_x h_0)^2$, which may account for the observed
behavior in the vicinity of zero.  Higher order terms may be
responsible for the complicated behavior observed for large tilt. }
\figure{ The convergence of the flow (3) to the
fluctuation-dissipation subspace for positive coefficients, starting
from different initial conditions.  For simplicity the symmetric coupling
is presented, i.e. $\nu_0 = \nu_1$, $D_0 = D_1$, $\lambda_0
= \lambda_1$, $\gamma_0 = \gamma_1$, $\varphi_0 = \varphi_1$ and
$\chi_0=\chi_1$.
 This choice  is motivated by symmetry between  $h_0(x,t)$ and $h_1(x,t)$
in the model.
Direct integration of (3) shows that the symmetric subspace is also
stable for small perturbations.
}
\figure{ Symmetric coupling: The scaling of the width
with time for systems
of size $L=50,100,200$ and $400$. An average over $50000, 10000, 5000$,
and $1500$ independent runs were taken.
  The main figure shows the scaling after an
intrinsic width [12] of magnitude 0.225 was extracted from the data. The
slope of the straight part gives $\beta_0=\beta_1= 0.32 \pm 0.01$. The
inset (a) shows the same data without extracting the intrinsic width.
The roughness exponent $\chi_0=0.52$ was determined from the best collapse of
the data, shown in inset (b). }
\figure{ The $\lambda_0=\varphi_i=\gamma_1=0$ case:
 The scaling of the width
with time for systems
of size $L=200,400$ and $800$. An average over $1000$ independent runs
were taken.
  The main figure shows the scaling after an
intrinsic width [12] of magnitude 0.225 was extracted from the data. The
slope of the straight part gives $\beta_0 = \beta_1= 0.33 \pm 0.02$.
The roughness exponent $\chi$ was determined from the best collapse of
the data, shown in inset (a). Inset (b) shows the scaling of the
height-height correlation function $< \vert h_i(x,t)-h_i(x+l,t) \vert^2>
\sim l^{2 \chi_i} $ for the interfaces $h_0(x,t)$ (empty symbols) and
$h_1(x,t)$ (filled symbols), for the same system sizes as in the main
picture. The slope of the straight line is
$2 \chi_0=1.28$. }

\begin{references}
\bibitem{[8]} Although the equations (1) are not always describing  the
evolution
of two surfaces, for   simplicity
we use the term {\it  interface } for $h_0(x,t)$ and  $h_1(x,t)$ throughout
this
paper.
\bibitem{[13]} EK [6] integrated the equations for  $\lambda_1 > 0$
and $\gamma_0 < 0$, but the exponents should coincide with those
obtained for
the opposite sign of the nonlinear terms:  $\lambda_1 < 0$ and
$\gamma_0 > 0$, since the dynamics is independent of a $h_i
\rightarrow -h_i$ change of variables.
\bibitem{[14]} We note that there is  no contradiction between  the possible
negativity of the $\gamma_i$ terms in the symmetric case and the
positivity in the  $\lambda_0=\varphi_i=\gamma_1=0$ case: the velocity
measurements also indicate a positive $\gamma_0$ term as a result of the
influence of $h_1$ on $h_0$, while in the symmetric case the negative
$\gamma_i$ may win in the competition of the two different signs.
\end{references}
 \end{document}